\documentclass[aps,prl,reprint,superscriptaddress]{revtex4-1}
\usepackage{blindtext}
\usepackage{xcolor}
\usepackage{hyperref}
\usepackage{epsfig}
\usepackage{subfigure}
\usepackage{graphicx}
\usepackage[toc,page]{appendix}
\usepackage{epstopdf}
\usepackage{ulem}
\usepackage{amsmath,amssymb,amsfonts}
\usepackage{array}
\usepackage{comment}
\usepackage{soul} 
\soulregister\cite7
\soulregister\ref7
\soulregister\pageref7
\usepackage[T1]{fontenc}

\usepackage{url}
\usepackage{hyperref}
\usepackage{lineno}
\usepackage{xspace}
\usepackage{listings}
\usepackage{float}
\usepackage{subfigure}
\usepackage{subfiles}

\graphicspath{{Figures/}{Figures_v2/}}

\begin{document}

\title{Non-Monotonicity of Transverse Momentum Correlations in Au + Au Collisions at RHIC}
\affiliation{Academia Sinica, Nankang, 115}
\affiliation{Panchayat College (PC), Bargarh, affiliated with Sambalpur University (SU), Odisha 768028, India}
\affiliation{Abilene Christian University, Abilene, Texas   79699}
\affiliation{AGH University of Krakow, FPACS, Cracow 30-059, Poland}
\affiliation{Argonne National Laboratory, Argonne, Illinois 60439}
\affiliation{American University in Cairo, New Cairo 11835, Egypt}
\affiliation{Ball State University, Muncie, Indiana, 47306}
\affiliation{Brookhaven National Laboratory, Upton, New York 11973}
\affiliation{University of Calabria \& INFN-Cosenza, Rende 87036, Italy}
\affiliation{University of California, Berkeley, California 94720}
\affiliation{University of California, Davis, California 95616}
\affiliation{University of California, Los Angeles, California 90095}
\affiliation{University of California, Riverside, California 92521}
\affiliation{Central China Normal University, Wuhan, Hubei 430079 }
\affiliation{University of Illinois at Chicago, Chicago, Illinois 60607}
\affiliation{Chongqing University, Chongqing, 401331}
\affiliation{Creighton University, Omaha, Nebraska 68178}
\affiliation{Czech Technical University in Prague, FNSPE, Prague 115 19, Czech Republic}
\affiliation{Technische Universit\"at Darmstadt, Darmstadt 64289, Germany}
\affiliation{National Institute of Technology Durgapur, Durgapur - 713209, India}
\affiliation{ELTE E\"otv\"os Lor\'and University, Budapest, Hungary H-1117}
\affiliation{Frankfurt Institute for Advanced Studies FIAS, Frankfurt 60438, Germany}
\affiliation{Fudan University, Shanghai, 200433 }
\affiliation{Guangxi Normal University, Guilin, 541004}
\affiliation{University of Heidelberg, Heidelberg 69120, Germany }
\affiliation{University of Houston, Houston, Texas 77204}
\affiliation{Huzhou University, Huzhou, Zhejiang  313000}
\affiliation{Indian Institute of Science Education and Research (IISER), Berhampur 760010 , India}
\affiliation{Indian Institute of Science Education and Research (IISER) Tirupati, Tirupati 517507, India}
\affiliation{Indian Institute Technology, Patna, Bihar 801106, India}
\affiliation{Indiana University, Bloomington, Indiana 47408}
\affiliation{Institute of Modern Physics, Chinese Academy of Sciences, Lanzhou, Gansu 730000 }
\affiliation{University of Jammu, Jammu 180001, India}
\affiliation{Kent State University, Kent, Ohio 44242}
\affiliation{University of Kentucky, Lexington, Kentucky 40506-0055}
\affiliation{Lanzhou University, Lanzhou, 730000}
\affiliation{Lawrence Berkeley National Laboratory, Berkeley, California 94720}
\affiliation{Lehigh University, Bethlehem, Pennsylvania 18015}
\affiliation{Lovely Professional University, Jalandhar - Delhi G.T. Road, Pagwara, Panjab, 144411, India}
\affiliation{Max-Planck-Institut f\"ur Physik, Munich 80805, Germany}
\affiliation{Michigan State University, East Lansing, Michigan 48824}
\affiliation{National Institute of Science Education and Research, HBNI, Jatni 752050, India}
\affiliation{National Cheng Kung University, Tainan 70101 }
\affiliation{Nuclear Physics Institute of the CAS, Rez 250 68, Czech Republic}
\affiliation{The Ohio State University, Columbus, Ohio 43210}
\affiliation{Panjab University, Chandigarh 160014, India}
\affiliation{Purdue University, West Lafayette, Indiana 47907}
\affiliation{Rice University, Houston, Texas 77251}
\affiliation{Rutgers University, Piscataway, New Jersey 08854}
\affiliation{University of Science and Technology of China, Hefei, Anhui 230026}
\affiliation{South China Normal University, Guangzhou, Guangdong 510631}
\affiliation{Sejong University, Seoul, 05006, Korea, Republic Of}
\affiliation{Shandong University, Qingdao, Shandong 266237}
\affiliation{Shanghai Institute of Applied Physics, Chinese Academy of Sciences, Shanghai 201800}
\affiliation{Southern Connecticut State University, New Haven, Connecticut 06515}
\affiliation{State University of New York, Stony Brook, New York 11794}
\affiliation{Instituto de Alta Investigaci\'on, Universidad de Tarapac\'a, Arica 1000000, Chile}
\affiliation{Temple University, Philadelphia, Pennsylvania 19122}
\affiliation{Texas A\&M University, College Station, Texas 77843}
\affiliation{Texas Southern University, Houston, Texas, 77004}
\affiliation{University of Texas, Austin, Texas 78712}
\affiliation{Tsinghua University, Beijing 100084}
\affiliation{University of Tsukuba, Tsukuba, Ibaraki 305-8571, Japan}
\affiliation{University of Chinese Academy of Sciences, Beijing, 101408}
\affiliation{United States Naval Academy, Annapolis, Maryland 21402}
\affiliation{Valparaiso University, Valparaiso, Indiana 46383}
\affiliation{Variable Energy Cyclotron Centre, Kolkata 700064, India}
\affiliation{Warsaw University of Technology, Warsaw 00-661, Poland}
\affiliation{Wayne State University, Detroit, Michigan 48201}
\affiliation{Wuhan University of Science and Technology, Wuhan, Hubei 430065}
\affiliation{Yale University, New Haven, Connecticut 06520}

\author{B.~E.~Aboona}\affiliation{Texas A\&M University, College Station, Texas 77843}
\author{J.~Adam}\affiliation{Czech Technical University in Prague, FNSPE, Prague 115 19, Czech Republic}
\author{L.~Adamczyk}\affiliation{AGH University of Krakow, FPACS, Cracow 30-059, Poland}
\author{I.~Aggarwal}\affiliation{Panjab University, Chandigarh 160014, India}
\author{M.~M.~Aggarwal}\affiliation{Panjab University, Chandigarh 160014, India}
\author{Z.~Ahammed}\affiliation{Variable Energy Cyclotron Centre, Kolkata 700064, India}
\author{A.~K.~Alshammri}\affiliation{Kent State University, Kent, Ohio 44242}
\author{E.~C.~Aschenauer}\affiliation{Brookhaven National Laboratory, Upton, New York 11973}
\author{S.~Aslam}\affiliation{Fudan University, Shanghai, 200433 }
\author{J.~Atchison}\affiliation{Abilene Christian University, Abilene, Texas   79699}
% ...

\collaboration{STAR Collaboration}\noaffiliation

\date{\today}

\begin{abstract}
Event-by-event transverse momentum correlations are sensitive to the equation of state of strongly interacting matter and are expected to exhibit   anomalous fluctuations in the vicinity of the QCD critical point. We report the first measurements of two-particle transverse momentum ($p_{\rm T}$) correlations for mid-rapidity charged particles in fixed-target Au+Au collisions at nucleon-nucleon center-of-mass energies $\sqrt{s_{\rm NN}} = 3.0$--$7.7$~GeV, measured by the STAR experiment during the Beam Energy Scan (BES) Phase~II program. The dependence of the scaled correlator on the number of participating nucleons ($N_{\text{part}}$) is studied to test expectations from an independent-source scenario, where the correlations are expected to scale as $1/\sqrt{N_{\text{part}}}$. We observe a clear breakdown of the expected scaling behavior in central collisions and identify a statistically significant non-monotonic dependence of the $p_{\rm T}$ correlations on collision energy, with a significance of approximately $5\sigma$. In contrast, transport-model calculations and data from mid-central collisions yield significances of only $2\sigma$ and $1.4\sigma$, respectively—insufficient to support a claim of non-monotonicity. These observations provide new constraints on the equation of state at high baryon density and may be sensitive to the presence of a QCD critical point.

\end{abstract}
\maketitle


\textbf{Introduction.} The goal of the Beam Energy Scan (BES) program at the Relativistic Heavy Ion Collider (RHIC) is to study the phase structure of nuclear matter governed by Quantum Chromodynamics (QCD) over a wide range of temperatures and baryon densities~\cite{STAR:2010vob,Bzdak_2020,Chen:2024aom}. Recently, a fixed-target (FXT) mode, implemented in the Solenoidal Tracker at RHIC (STAR) experiment, has extended the coverage to even higher baryon densities, up to a baryon chemical potential ($\mu_{B}$) of approximately 760 MeV~\cite{STAR:2020dav}. This region is close to the conjectured critical point from recent theoretical calculations near $\mu_{B} = 600$ MeV~\cite{PhysRevD.101.054032,PhysRevD.104.054022,db22d315623f47ebb2d44d269787395d,PhysRevC.110.015203,clarke2024searchingqcdcriticalendpoint,shah2024locatingqcdcriticalpoint,sorensen2024locatingcriticalpointhadron}. The observation of non-monotonic behavior in event-by-event correlations and fluctuations of global quantities is often quoted as a key search strategy for the QCD critical point~\cite{PhysRevLett.85.2689,PhysRevC.60.024901,Heiselberg_2001,Koch:2025cog}. In the past, measurements of particle multiplicity fluctuations have mainly been employed to probe critical phenomena~\cite{PhysRevD.60.114028,Pradeep:2025uej,STAR:2010mib,STAR:2013gus,STAR:2020tga,9l69-2d7p,STAR:2020tga}, but transverse-momentum correlations have also been proposed, not only as a measure of thermalization~\cite{PhysRevLett.92.162301,PhysRevC.85.014905}, but also as a probe for the critical point~\cite{PhysRevD.65.096008}. As is the case for any kinematic observable, transverse-momentum fluctuations are inherently susceptible to non-critical dynamical backgrounds~\cite{Du:2025tdh,JETSCAPE:2020mzn}. However, the extent to which these non-critical backgrounds alone can account for the observed non-monotonic beam-energy dependence remains to be established.

Dynamical fluctuations of the mean transverse momentum have been studied in nucleus–nucleus (A–A) collisions at the Super Proton Synchrotron (SPS), at RHIC~\cite{STAR:2019dow,STAR:2024wgy,PHENIX:2003ccl,STAR:2025elk}, and at the LHC~\cite{ALICE:2014gvd,ATLAS:2024jvf}. These measurements are expected to offer access to temperature and energy fluctuations~\cite{PhysRevLett.75.1044,Shuryak_1998,Cao:2021zhy} and, arguably, their magnitude is proportional to the heat capacity of the created hot QCD system. The particular form of the temperature dependence of the heat capacity in the vicinity of the QCD phase transition depends on the order of the transition~\cite{Chen:2025vwl}. At a possible critical point, which marks the onset of a first-order phase transition, the heat capacity is expected to diverge; that is, temperature fluctuations should then be strongly suppressed~\cite{Shuryak_1998,jeon2003eventbyeventfluctuations,Basu_2016}.

Crucially, previous measurements of mean transverse momentum fluctuations were conducted exclusively in collider mode and focused on the lower-$\mu_{\rm B}$ region of the QCD phase diagram. In contrast, this paper reports the first measurement of dynamical fluctuations of the mean transverse momentum in the high-baryon-density regime accessed through the STAR FXT mode with upgraded detectors, providing essential constraints on the properties of QCD matter in proximity to the anticipated critical point. In addition to their sensitivity to critical fluctuations, transverse momentum correlations probe the early-time pressure gradients in the medium. In hydrodynamic frameworks, these correlations are directly linked to the speed of sound, $c_{\rm s}$~\cite{Gardim_2024,skhj-cj9p}, and therefore provide sensitivity to the equation of state of QCD matter~\cite{Sorensen:2023zkk,liu2025newconstraintsequationstate}.

In this Letter, we present an experimental study of the collision-energy dependence of $p_{\rm T}$ correlations ($C_{p_{\rm T}}$) using Au+Au collisions at nucleon–nucleon center-of-mass energies of $\sqrt{s_{\rm NN}} = 3.0$–$7.7$ GeV in the high-baryon-density region ($\mu_B \approx 760$–$400$ MeV), taken during the RHIC BES II program. We report the first systematic measurement of $C_{p_{\rm T}}$ in the $\mu_B > 400$ MeV region and observe a clear breakdown of the expected $N_{\text{part}}$-scaling behavior. Most importantly, a significant non-monotonic structure in the energy dependence of $p_{\rm T}$ correlations for central collisions is identified, providing new, stringent constraints on the equation of state at high baryon densities.
\begin{figure*}[!htbp]
	\centering
	\includegraphics[scale=0.8]{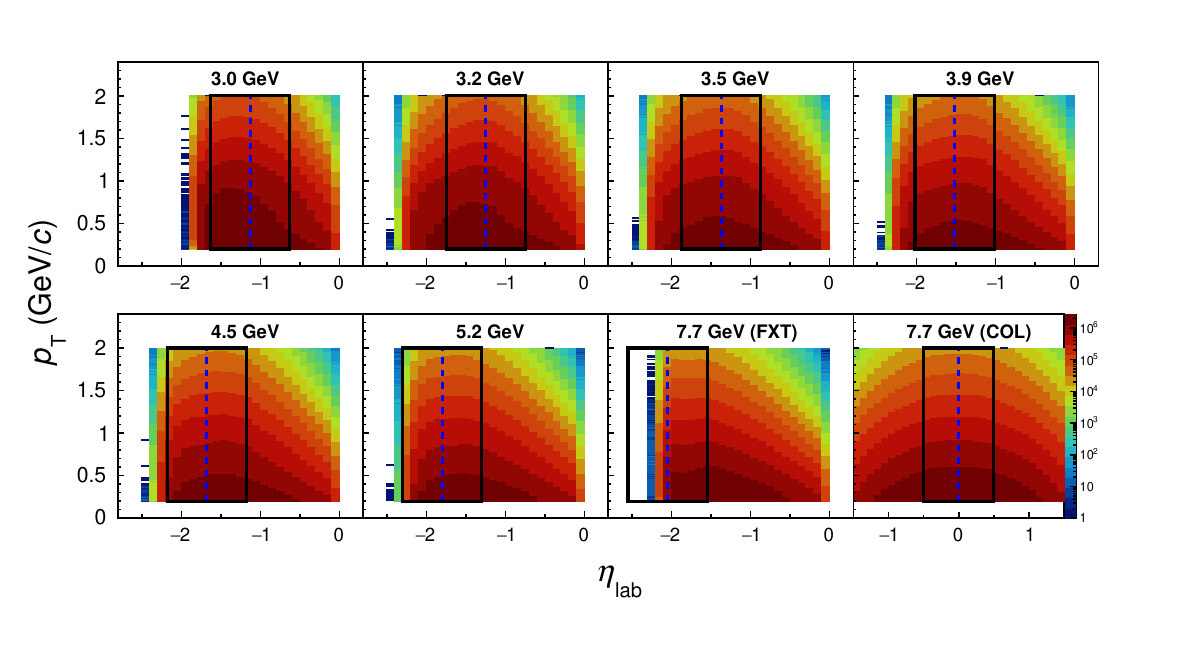}
	\caption{\label{fig:Phase space}Efficiency-uncorrected density distributions in transverse momentum ($p_{\rm T}$) and particle pseudorapidity ($\eta_{\rm lab}$) in the lab frame for 0--5$\%$ centrality Au+Au collisions measured with the STAR TPC at $\sqrt{s_{\rm NN}} = 3.0$, 3.2, 3.5, 3.9, 4.5, 5.2, 7.7 (FXT) and 7.7 (COL) GeV. For $\sqrt{s_{\rm NN}} = 7.7$ GeV, the collider-mode acceptance is shown to illustrate the difference between fixed-target and collider configurations. The dashed blue lines indicate midrapidity, and the black box marks the analysis window. Each panel is self-normalized along the z axis.}
\end{figure*}
\newline \textbf{Analysis.} The primary FXT event selection criteria were defined to ensure that collisions occurred within the Au target volume and to reject background events. The primary vertex position along the beam direction ($V_{z}$) was required to be within $2\,{\rm cm}$ of the target center, which is located $200.7\,{\rm cm}$ from the center of the Time Projection Chamber (TPC)~\cite{STAR:2002eio}. To exclude background events generated in the vacuum pipe, the primary vertex radial position ($V_{r}$) was constrained to be less than $2\,{\rm cm}$. For $\sqrt{s_{\rm NN}} = 3.0$ GeV, only the TPC was used, covering the laboratory-frame pseudorapidity range $-2.0 < \eta_{\rm lab} < 0$. For higher fixed-target energies ($\sqrt{s_{\rm NN}} =$ 3.2, 3.5, 3.9, 4.5, 5.2, and 7.7 GeV), the TPC combined with the inner TPC (iTPC) upgrade~\cite{Shen:2018pkc} extended the coverage to $-2.4 < \eta_{\rm lab} < 0$. Additionally, 7.7 GeV collider-mode data were analyzed within $-1 < \eta < 1$ in the center-of-mass frame.

The final dataset used for this analysis comprises the following approximate event counts across the BES-II program: $260\,\mathrm{M}$ ($3.0\,\mathrm{GeV}$), $200\,\mathrm{M}$ ($3.2\,\mathrm{GeV}$), $120\,\mathrm{M}$ ($3.5\,\mathrm{GeV}$), $80\,\mathrm{M}$ ($3.9\,\mathrm{GeV}$), $86\,\mathrm{M}$ ($4.5\,\mathrm{GeV}$), $100\,\mathrm{M}$ ($5.2\,\mathrm{GeV}$), $50\,\mathrm{M}$ ($7.7\,\mathrm{GeV}$, fixed target), and $126\,\mathrm{M}$ ($7.7\,\mathrm{GeV}$, collider).

Collision centrality is determined from charged primary tracks reconstructed in the TPC. For fixed-target data, centrality is defined without an explicit pseudorapidity requirement; the effective acceptance is approximately $-2.0 < \eta_{\rm lab} < 0$ for the TPC alone and extends to $-2.24 < \eta_{\rm lab} < 0$ when the iTPC is available. For collider data, centrality is constructed using tracks within $|\eta| < 0.5$ and $p_{\rm T} > 0.15$ GeV/$c$. The centrality calibration is corroborated using Monte Carlo Glauber simulations~\cite{Miller_2007}, with corrections applied to account for collision vertex position and the luminosity dependence.

The main detectors used were the TPC~\cite{STAR:2002eio} and the Time-of-Flight (TOF) detector~\cite{Llope:2012zz}, both located in a solenoidal magnetic field of 0.50~T. Charged tracks reconstructed in the TPC with $0.2 \leq p_{\rm T} \leq 2.0~\mathrm{GeV}/c$ and $|\eta_{\rm cm}| < 0.5$ were used in this analysis. Here, $\eta_{\rm cm} = \eta_{\rm lab} - \eta_{\rm mid}$ accounts for the shift of midrapidity in the fixed-target mode, where $\eta_{\rm mid}$ varies with $\sqrt{s_{\rm NN}}$. Tracks in the TPC were characterized by the distance of closest approach (DCA), defined as the smallest distance between the extrapolated track and the measured event vertex. Each track was required to have at least 15 measured space points. Each event was required to have at least two tracks matched to TOF hits to minimize pileup. Statistical uncertainties were estimated using the bootstrap method~\cite{Luo:2014rea}.
Systematic uncertainties were evaluated by varying key analysis cuts and selections. We varied the distance of closest approach (DCA) cut from 2.8 to 3.2 cm, the pseudorapidity acceptance from $|\eta_{\rm cm}| < 0.45$ to $< 0.55$, the minimum number of hit points required for track reconstruction (NhitsFit) to 12, and the lower $p_{\rm T}$ cut from 0.18 to 0.22~GeV/$c$. In addition, two-dimensional closure tests in $(p_{\rm T},\eta)$ were performed using acceptance and efficiency maps from STAR embedding~\cite{FINE200176} for fixed-target data. The observable calculated after applying detector acceptance and efficiency effects was consistent with the true value within statistical uncertainties, confirming that the observable is insensitive to efficiency corrections. The resulting closure uncertainty when extrapolating from fixed-target to collider energies is estimated to be approximately 2--3\% (relative); to preserve direct comparability with previously published STAR measurements, this effect is not quoted as a separate systematic contribution. Barlow checks~\cite{barlow2002systematicerrorsfactsfictions} confirmed that the observed variations are significant relative to statistical uncertainties. For the most central collisions, the average relative systematic uncertainties associated with DCA, NhitsFit, $\eta$ acceptance, and lower $p_{\rm T}$ cut are 0.03\%, 0.04\%, 0.07\%, and 0.18\%, respectively. Overall, systematic uncertainties dominate over statistical ones and are shown as gray shading in the figures.

\textbf{Observables.} Fluctuations of the event-by-event mean transverse momentum, $\langle p_{\rm T} \rangle$, of charged particles are investigated using two-particle $p_{\rm T}$ correlators. The event-by-event mean transverse momentum is defined as
\begin{equation}
\langle p_{\rm T} \rangle = \frac{\sum_{i=1}^{N_{\rm ch}} p_{{\rm T},i}}{N_{\rm ch}},
\end{equation}
where $p_{{\rm T},i}$ is the transverse momentum of the $i$th particle and $N_{\rm ch}$ is the total number of charged particles in the event. Alternatively, the standard moment method can be employed for the event-by-event analysis of $\langle p_{\rm T} \rangle$ fluctuations. This method calculates moments of the $\langle p_{\rm T} \rangle$ distribution, providing a comprehensive evaluation of the total fluctuation by accounting for both statistical and dynamical (non-statistical) contributions. 

In contrast, multiparticle $p_{\rm T}$ correlators are sensitive only to dynamical fluctuations and therefore provide a direct measure of the correlations of interest. This feature makes $p_{\rm T}$ correlations comparatively robust against volume fluctuations and minimally sensitive to the details of the centrality selection, as supported by the consistency observed between the 7.7 GeV collider and fixed-target measurements despite their differing phase-space coverage and centrality definitions (see Fig.3 of the Supplemental Material~\cite{Supplemental} \nocite{Luo_2013} and Ref.~\cite{Kovalenko_2017}). The expression for the two-particle $p_{\rm T}$ correlator, $\langle \Delta p_{{\rm T},i} \Delta p_{{\rm T},j} \rangle$, is given by

\begin{align} \label{eq3}
\langle \Delta p_{{\rm T},i} \Delta p_{{\rm T},j} \rangle 
&= \left\langle \frac{\sum\limits_{i,j,\,i\neq j}^{N_{\rm ch}} (p_{{\rm T},i} - \langle\langle p_{\rm T} \rangle\rangle)(p_{{\rm T},j} - \langle\langle p_{\rm T} \rangle\rangle)}{N_{\rm ch}(N_{\rm ch} - 1)} \right\rangle_{\rm ev} \notag \\
&= \left\langle \frac{Q_1^2 - Q_2}{N_{\rm ch}(N_{\rm ch} - 1)} \right\rangle_{\rm ev}
- \left\langle \frac{Q_1}{N_{\rm ch}} \right\rangle_{\rm ev}^2 ,
\end{align}
where $Q_n = \sum_{i=1}^{N_{\rm ch}} (p_{{\rm T},i})^n$. Since $Q_1^2$ contains contributions from both distinct particle pairs ($i \neq j$) and diagonal self-pairs ($i=j$), the subtraction of $Q_2$ removes the self-correlation terms and reproduces the original pair sum restricted to $i \neq j$. 
This well-established $Q$-vector framework~\cite{Acharya_2024,PhysRevC.105.024904,PhysRevC.105.014906,Kitazawa:2025xcy} also avoids the computational cost of nested loops. The terms in Eq.~\ref{eq3} are evaluated using single loops over particles per event. Here, $\langle \cdots \rangle_{\rm ev}$ denotes an event-ensemble average and $\langle\langle p_{\rm T} \rangle\rangle = \langle Q_1 / N_{\rm ch} \rangle_{\rm ev}$ is the event-averaged mean transverse momentum.

The final observable presented in this Letter is the relative dynamical correlation,
\begin{equation}
C_{p_{\rm T}} \equiv 
\frac{\sqrt{\langle \Delta p_{{\rm T},i}\Delta p_{{\rm T},j} \rangle}}
{\langle\langle p_{\rm T}\rangle\rangle}.
\end{equation}
This two-particle transverse-momentum correlator is directly related to the collective radial-flow fluctuation $v_0$~\cite{osti_1646590,ATLAS:2025ztg,ALICE:2025iud}.The variance of $\langle p_{\rm T} \rangle$ can be decomposed into a purely statistical term and a dynamical contribution proportional to $\langle \Delta p_{{\rm T},i}\Delta p_{{\rm T},j} \rangle$, which reflects genuine two-particle correlations.

The dimensionless form of $C_{p_{\rm T}}$ enables direct comparison with previous measurements by the CERES~\cite{CERES:2008wlj} and ALICE~\cite{ALICE:2014gvd} Collaborations and offers several practical advantages, including reduced sensitivity to momentum-scale uncertainties, partial independence from the absolute value of $\langle p_{\rm T} \rangle$~\cite{ALICE:2014gvd}, and  reduced sensitivity to collective radial flow effects, since the flow-induced enhancement in the numerator is largely canceled by the corresponding increase in the average transverse momentum in the denominator~\cite{Abdel-Aziz:2005dan}.

\begin{figure}[htbp!]
\includegraphics[width=\linewidth]{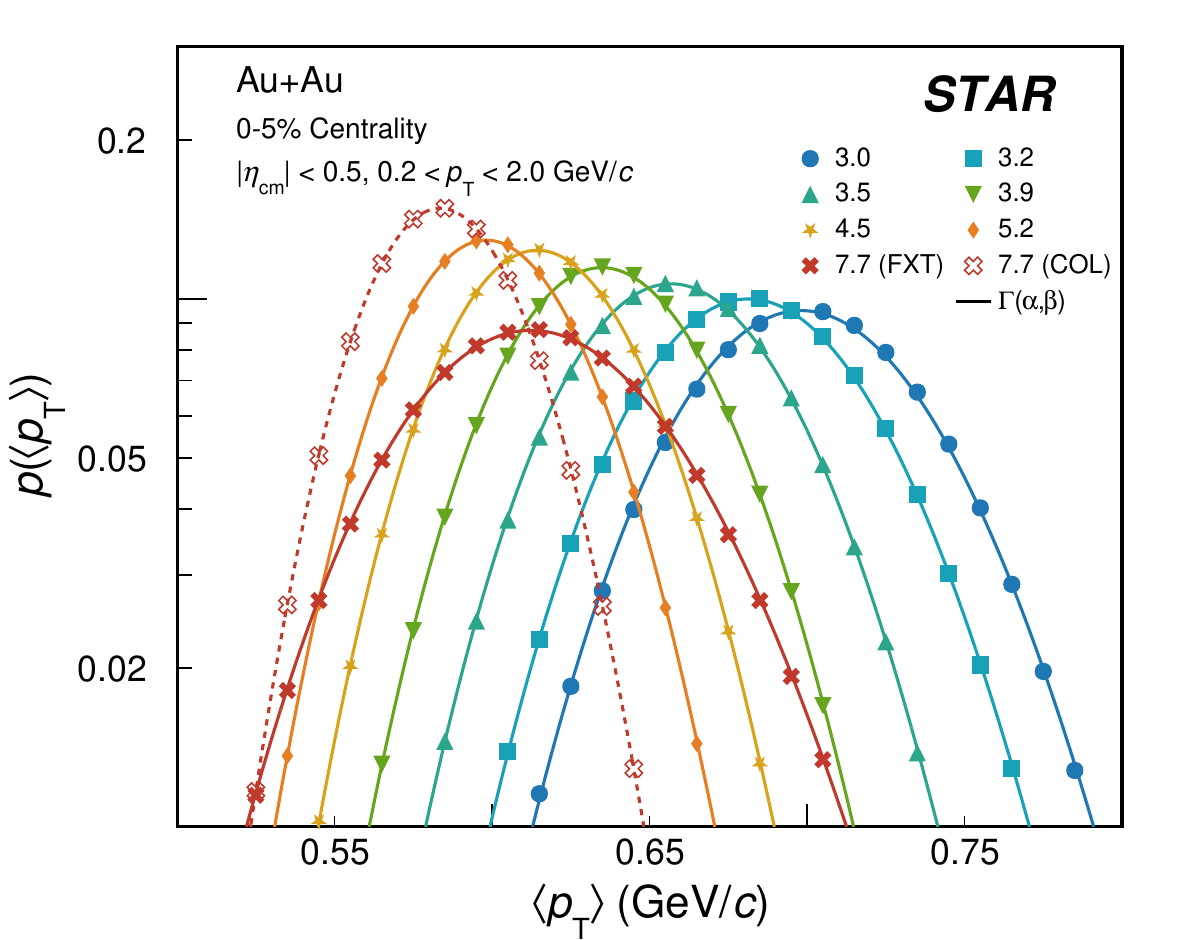}
\caption{\label{fig:Mean}Histograms of the uncorrected average transverse momentum per event at mid-pseudorapidity ($|\eta_{\rm cm}| < 0.5$) and for $0.2 < p_{\rm T} < 2.0$ GeV/$c$ in Au+Au collisions at $\sqrt{s_{\rm NN}} = 3.0$, 3.2, 3.5, 3.9, 4.5, 5.2, 7.7 (FXT) and 7.7 (COL) GeV for the top 5\% most central collisions at each energy. The solid lines represent Gamma distribution fits.}
\end{figure}

\begin{figure*}[htbp]
\includegraphics[scale=0.8]{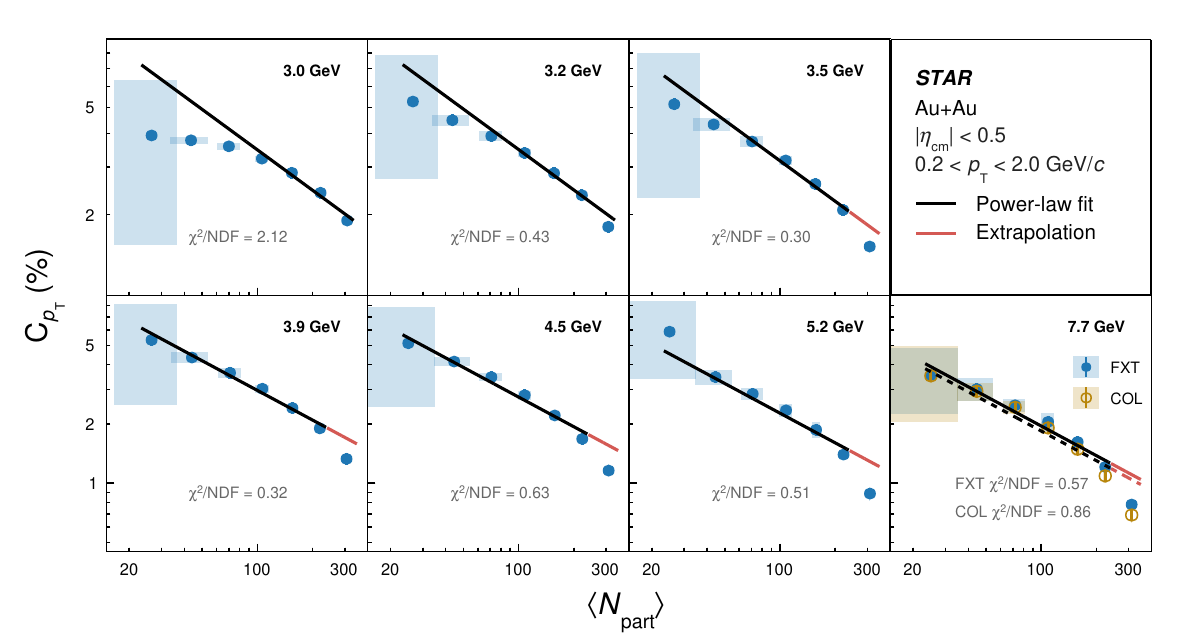}
\caption{\label{fig:Npart}The relative dynamical transverse momentum correlation \(C_{p_{\rm T}}\) as a function of the number of participating nucleons \(\langle N_{\rm part}\rangle\) in Au+Au collisions at several collision energies. Results are shown for \(\sqrt{s_{\rm NN}} = 3.0\), 3.2, 3.5, 3.9, 4.5, 5.2, and 7.7~GeV. The 7.7~GeV data include both Fixed-Target (FXT, blue circles) and Collider (COL, yellow circles) configurations. Statistical uncertainties are shown as vertical bars and systematic uncertainties as shaded boxes. The black lines represent power-law fits with the normalization determined from the data. The red lines show the extrapolation of these fits to lower \(\langle N_{\rm part}\rangle\). Data points correspond to 10\%-wide centrality bins. Charged particles are selected within \(0.2 < p_{\rm T} < 2.0\) GeV/$c$ and \(|\eta_{\rm cm}| < 0.5\).}
\end{figure*}


\textbf{Results.} Figure~\ref{fig:Phase space} illustrates the charged-particle distributions in the $p_{\rm T}$--$\eta_{\rm lab}$ phase space for the energies studied: $\sqrt{s_{\rm NN}} = 3.0, 3.2, 3.5, 3.9, 4.5, 5.2, 7.7$ (FXT), and $7.7$ (collider) GeV. In the fixed-target (FXT) configuration, midrapidity ($\eta_{\rm cm} = 0$) shifts toward negative $\eta_{\rm lab}$ values as the collision energy increases. This is in contrast to the collider mode, where $\eta_{\rm cm} = 0$ is fixed near the center of the detector ($\eta_{\rm lab} \approx 0$). The analysis window ($|\eta_{\rm cm}| < 0.5$) is indicated by a black box, with the mid-pseudorapidity at each energy marked by a blue dashed line.

For the lower FXT energies, the analysis window is well-contained within the TPC volume. However, at $\sqrt{s_{\rm NN}} = 7.7$ GeV (FXT), the center-of-mass frame shift begins to push the window toward the detector boundary, resulting in a slight acceptance loss at the edges of the TPC. To ensure that our physics conclusions are not biased by these edge effects, we incorporate a complementary measurement from 7.7 GeV collider data. The collider setup provides full, symmetric acceptance at the same center-of-mass energy  within $|\eta_{\rm cm}| < 0.5$. The consistency observed between the FXT and collider results at 7.7 GeV validates that the slight acceptance limitation in the FXT mode does not introduce significant systematic biases, ensuring the robustness of the $C_{p_T}$ observable across the entire energy range.

A key consideration in critical-point searches is minimizing detector-dependent correlations and acceptance effects between multiplicity determination and observable measurements. Using the same detector subsystem, such as the TPC, provides a consistent acceptance and reconstruction framework for these measurements. This approach minimizes detector-induced fluctuations and ensures uniformity across events, thereby enhancing the robustness of fluctuation and correlation measurements~\cite{PhysRevC.111.034902}.(The TOF detector contributes solely to pileup suppression via event selection criteria.)

Event-by-event $\langle p_{\rm T} \rangle$ distributions for Au+Au collisions at $\sqrt{s_{\rm NN}} = 3.0$, 3.2, 3.5, 3.9, 4.5, 5.2, and 7.7 GeV for the top 5$\%$ centrality are shown in Fig.~\ref{fig:Mean}. The distributions are normalized by the total number of events and are not corrected for detector efficiencies. They are fitted with a Gamma function ($\Gamma$)~\cite{Tannenbaum:2001gs}. We observe that both the mean and the width of the distributions, corresponding to the first and second order moments, exhibit a monotonic energy dependence over the studied range, increasing as $\sqrt{s_{\rm NN}}$ decreases. This behavior is consistent with enhanced contributions from primordial protons, which shift the mean to higher values, and with reduced particle multiplicity at lower energies, which leads to a broadening of the distributions.

The event-by-event $\langle p_{\rm T} \rangle$ distributions differ between the fixed-target and collider configurations at $\sqrt{s_{\rm NN}} = 7.7$~GeV, as shown in Fig.~\ref{fig:Mean}. This difference is expected because $\langle p_{\rm T} \rangle$ is an extensive quantity and is therefore sensitive to phase-space losses and detector efficiency differences between the two modes. In contrast, the dynamical correlation quantified by $C_{p_{\rm T}}$ is designed to be insensitive to such undesired acceptance and efficiency effects. As demonstrated in Fig.~\ref{fig:Npart} by the consistent results obtained in both configurations, the correlator isolates genuine dynamical fluctuations. This consistency is quantified for each centrality bin using both UrQMD simulations and STAR data (see Figs.1 and 2 in the Supplemental Material~\cite{Supplemental}).

In Figure~\ref{fig:Npart}, the relative dynamical correlation $C_{p_{\rm T}}$ is shown as a function of centrality for several collision energies. 
The correlations are observed to decrease towards central collisions.

To assess whether the observed transverse-momentum fluctuations reflect purely statistical behavior or contain contributions from critical dynamics, we examine the centrality dependence of $C_{p_{\rm T}}$ shown in Fig.~\ref{fig:Npart}. In the absence of collective or critical effects, the correlations are expected to follow a simple statistical baseline arising from the superposition of many independent particle-emitting sources. This scenario leads to an approximate power-law dependence of the form $A(\sqrt{s_{\rm NN}})/\sqrt{N_{\rm part}}$, reflecting the dilution of fluctuations with increasing system size. Any deviation from this scaling therefore provides a sensitive probe of non-trivial dynamics, including possible modifications to the effective degrees of freedom in the strongly interacting QCD medium or density-dependent effects, which may be enhanced in the most central collisions and could reflect proximity to a critical point~\cite{PhysRevC.92.024915,PhysRevLett.92.162301}.

At the lowest energy, $\sqrt{s_{\rm NN}} = 3.0$~GeV, the system is strongly dominated by hadronic interactions, and the independent-source picture breaks down entirely; indeed, no meaningful power-law behavior can be established at this energy. Beginning at $\sqrt{s_{\rm NN}} = 3.2$~GeV, however, the statistical hypothesis becomes applicable, and power-law fits can be performed on the data. Because the available multiplicity at these energies does not permit high granularity in centrality, we follow a constrained procedure in which the power-law exponent is fixed to the expected baseline value of $-0.5$. This differs from analyses at higher energies at the LHC~\cite{ALICE:2014gvd,ATLAS:2024jvf}, where the exponent can be treated as a free parameter due to higher multiplicity and better centrality resolution.  

Even under this conservative treatment, we observe a clear and increasing deviation from the expected scaling in the most central bin as $\sqrt{s_{\rm NN}}$ rises from 3.2 to 7.7~GeV, suggesting the emergence of additional dynamical contributions beyond trivial statistical fluctuations~\cite{PhysRevLett.92.162301}.

\begin{figure}[htbp]
\includegraphics[width=\linewidth]{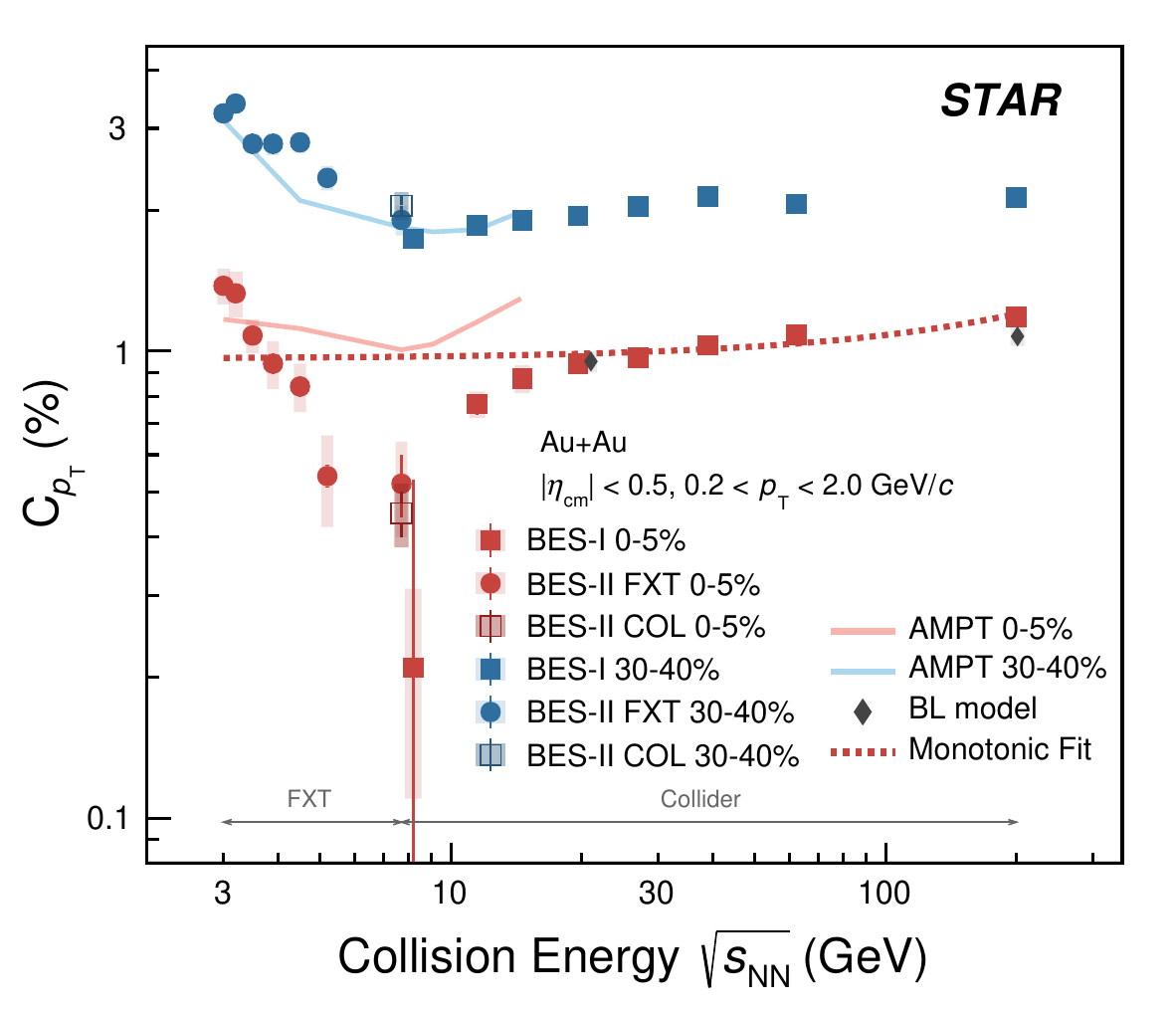}
\caption{\label{fig:Corr}The relative dynamical transverse momentum correlation \(C_{p_{\rm T}}\) as a function of collision energy \(\sqrt{s_{\rm NN}}\) in Au+Au collisions. Results are shown for 0--5\% (red markers) and 30--40\% (blue markers) centralities from BES-I, BES-II Fixed-Target (FXT), and BES-II Collider (COL) datasets. The newly measured BES-II collider point at \(\sqrt{s_{\rm NN}} = 7.7\) GeV is shown as a hollow square. Statistical uncertainties are indicated by vertical bars and systematic uncertainties by shaded boxes. The \(\sqrt{s_{\rm NN}} = 7.7\) GeV BES-I point is slightly shifted horizontally for visibility. Charged particles are selected within $0.2 < p_{\rm T} < 2.0$)~GeV/$c$ and $(|\eta_{\rm cm}| < 0.5$. AMPT~\cite{PhysRevC.111.024911} (solid lines) are shown for both centralities. Boltzmann-Langevin (BL) model~\cite{PhysRevC.95.064901} predictions are displayed as black diamonds at \(\sqrt{s_{\rm NN}} = 19.6\) and \(200\) GeV. A monotonic fit to the 0--5\% data is shown as a dotted red line.}
\end{figure}


Figure~\ref{fig:Corr} shows the relative dynamical correlation $C_{p_{\rm T}}$ as a function of $\sqrt{s_{\rm NN}}$ for the most central bin (0–5\%) along with previous STAR measurements from BES~I~\cite{STAR:2019dow}, allowing a direct comparison over a wide range of beam energies. These data are compared against calculations from the A Multi-Phase Transport (AMPT)~\cite{Lin:2004en,PhysRevC.111.024911} and Boltzmann–Langevin (BL)~\cite{PhysRevC.95.064901} models.

Although the AMPT model is known to describe the bulk $p_{\rm T}$ spectra reasonably well across these collision energies, it fails to reproduce the observed $p_{\rm T}$ correlations~\cite{PhysRevC.111.024911}, in particular the non-monotonic dependence on collision energy seen in the data. In 0–5$\%$ central collisions, the model qualitatively follows the overall energy dependence but fails to capture the non-monotonic structure in the intermediate-energy region. In contrast, the 30–40$\%$ centrality class is comparatively well described and exhibits a much weaker energy dependence. In addition, we compare our results to calculations based on the Boltzmann–Langevin approach, which incorporates fluctuation–dissipation dynamics relevant for equilibration and thermalization. These calculations show minimal dependence on the collision energy and are consistent with the STAR measurements at $\sqrt{s_{\rm NN}} = 19.6$ and 200~GeV~\cite{PhysRevC.95.064901}.  

The new high-statistics data at $\sqrt{s_{\rm NN}} = 7.7$~GeV significantly reduce uncertainties compared to BES~I and remain fully consistent with previous measurements. To quantify deviations from a smooth energy dependence, the significance of the observed non-monotonicity is evaluated by fitting the data to a strictly monotonic baseline, chosen as a first-order polynomial anchored to the collider-energy points, which carry the smallest statistical uncertainties (20-200 GeV). The deviation of the measured values from this polynomial reference is then used to define the statistical significance of the non-monotonic behavior.  

Using this procedure, a significance of approximately $5\sigma$ is obtained for the STAR 0--5\% central data (see Fig.4 of the Supplemental Material~\cite{Supplemental}). In contrast, the AMPT model calculations, analyzed under identical conditions and using the same methodology and uncertainty as the data, exhibit only a $\sim 1.4\sigma$ deviation, which is insufficient to support even weak evidence for non-monotonic behavior.

Applying the same analysis technique to mid-central collisions, a deviation of $\sim 2\sigma$ is observed in the 30–40\% centrality class. While this level of significance does not constitute conclusive evidence for non-monotonic behavior, it is consistent with the trend observed in the most central collisions and suggests a possible centrality dependence of the effect. This data-driven approach provides a robust and uniform framework for comparisons across centrality and is expected to motivate further theoretical and model investigations.

\textbf{Summary.} 
We have measured the relative dynamical correlation, $C_{p_{\rm T}}$, in Au+Au collisions at $\sqrt{s_{\rm NN}} = 3$–$7.7$~GeV, corresponding to high baryon densities ($\mu_B \approx 760$–$400$~MeV), along with event-by-event $\langle p_{\rm T} \rangle$ distributions. The data exhibit an approximate power-law scaling with $\langle N_{\rm part} \rangle$ for most collision energies, consistent with expectations from independent-source models and similar to observations in high-energy RHIC and LHC measurements. This scaling, however, breaks down at the lowest beam energy, where the system is expected to be hadronic throughout the collision process.  

More importantly, in central collisions, the correlations show a non-monotonic dependence on collision energy with a significance of approximately $5\sigma$. The observed features are robust with respect to the choice of reference frame: the measurements are performed at mid-pseudorapidity in the center-of-mass frame and remain consistent between fixed-target and collider configurations. In addition, the two-particle $p_{\rm T}$ correlator is designed to be insensitive to volume fluctuations and centrality bin-width corrections (CBWC)~\cite{Luo:2014rea}, ensuring that the non-monotonicity reflects genuine dynamical behavior rather than trivial geometric effects (see Fig.3 in the Supplemental Material~\cite{Supplemental}).  

Theoretical calculations place the QCD critical point within this same $\sqrt{s_{\rm NN}}$ range~\cite{shah2024locatingqcdcriticalpoint,borsanyi2025latticeqcdconstraintscritical} which overlaps with the energy region where the most pronounced dip is observed in 0–5$\%$ central collisions. These results therefore indicate the presence of non-trivial dynamical correlations in strongly interacting matter and suggest that the observed energy dependence may be influenced by critical phenomena, thereby providing new constraints on the possible location of the conjectured QCD critical point. Further theoretical studies incorporating critical dynamics will be essential for a full interpretation of the observed behavior. Recent BES-II net-proton measurements have reported a minimum in the ratio $C_4/C_2$ at $\sqrt{s_{\rm NN}} = 19.6$ GeV, where significant deviations from non-critical model baselines are observed in central Au+Au collisions~\cite{STAR:2020tga,9l69-2d7p}. Together, these complementary observables provide valuable constraints on possible critical dynamics in the QCD phase diagram, although a quantitative connection between different fluctuation measures remains under active investigation. Future measurements of higher-order $p_{\rm T}$ cumulants, along with extended net-proton cumulant data at lower energies, will be important for a more complete understanding.

\textbf{Acknowledgment.} 
We thank the RHIC Operations Group and SCDF at BNL, the NERSC Center at LBNL, and the Open Science Grid consortium for providing resources and support.  This work was supported in part by the Office of Nuclear Physics within the U.S. DOE Office of Science, the U.S. National Science Foundation, National Natural Science Foundation of China, Chinese Academy of Science, the Ministry of Science and Technology of China and the Chinese Ministry of Education, NSTC Taipei, the National Research Foundation of Korea, Czech Science Foundation and Ministry of Education, Youth and Sports of the Czech Republic, Hungarian National Research, Development and Innovation Office, New National Excellency Programme of the Hungarian Ministry of Human Capacities, Department of Atomic Energy and Department of Science and Technology of the Government of India, the National Science Centre and WUT ID-UB of Poland, German Bundesministerium f\"ur Bildung, Wissenschaft, Forschung and Technologie (BMBF), Helmholtz Association, Ministry of Education, Culture, Sports, Science, and Technology (MEXT), Japan Society for the Promotion of Science (JSPS), and Agencia Nacional de Investigacion y Desarrollo de Chile (ANID), Chile.

\bibliographystyle{apsrev4-1} 
\bibliography{apssamp}
\clearpage

\section*{Supplemental Material}
\setcounter{figure}{0}
\subsection{Effect of Frame of Reference on Measurement}
\label{sec:supplemental-frame}
In this Letter, we have reported measurements for Au+Au collisions at $\sqrt{s_{NN}} = 7.7$~GeV in both collider and fixed-target modes. This section presents cross-checks performed to ensure that the correlator measurements are independent of the chosen reference frame, allowing for a consistent comparison between the two configurations.
\begin{figure}[h!]
\includegraphics[width=0.95\linewidth]{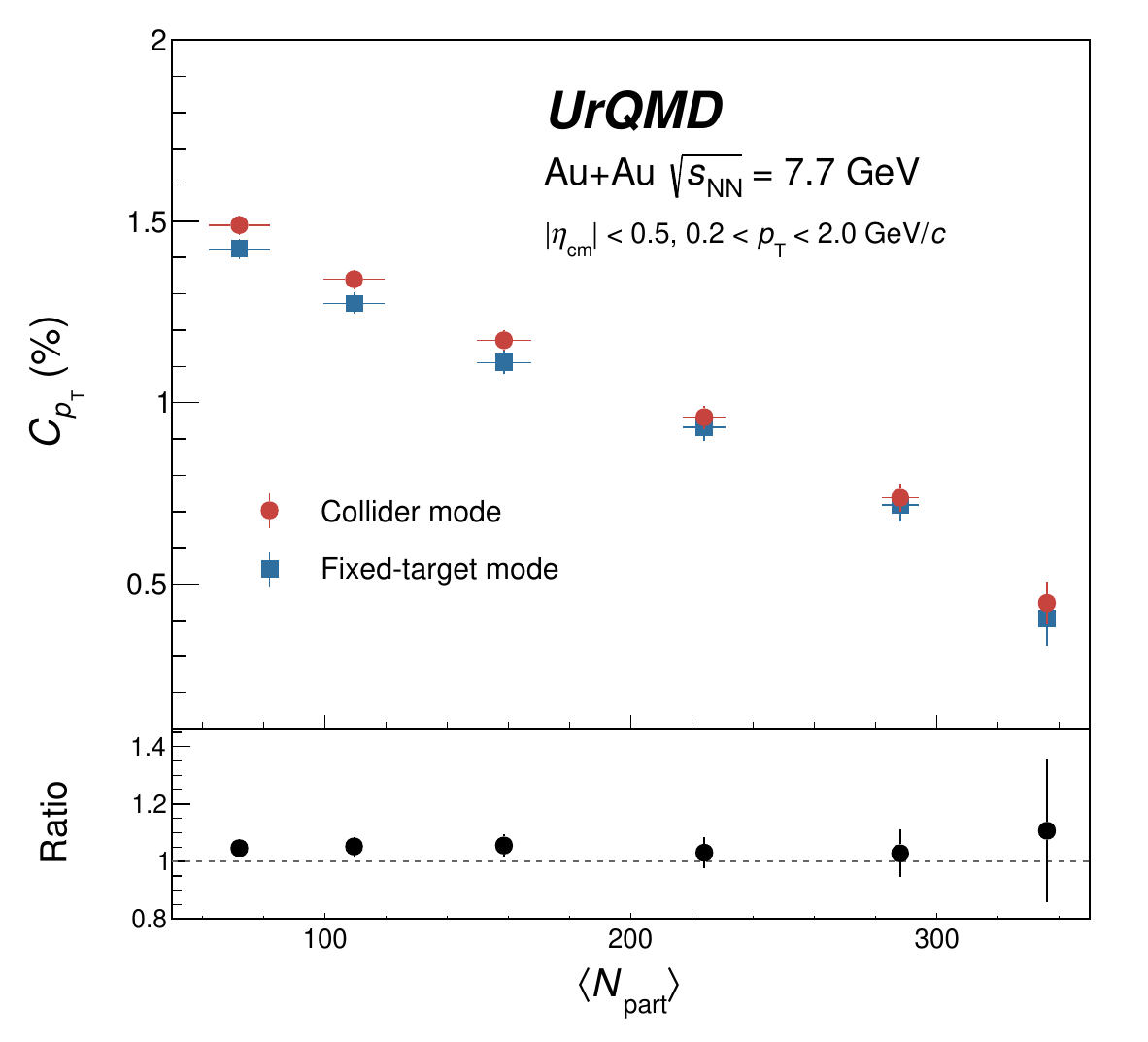}
\caption{\label{fig:S1}
The relative dynamical correlation $C_{p_{\rm T}}$ as a function of $\langle N_{\text{part}}\rangle$ for Au+Au collisions at $\sqrt{s_{\rm NN}} = 7.7$~GeV in collider and fixed-target modes, calculated using UrQMD ver.~3.3. The lower panel shows the ratio of collider-mode to fixed-target-mode results, demonstrating consistency between the two configurations within uncertainties.
}
\end{figure}

UrQMD events are generated in the collider frame (center-of-mass). To emulate a fixed-target configuration, each particle is boosted from the center-of-mass frame to the laboratory frame. Since rapidity is an additive quantity, the boost is applied in rapidity space, after which pseudorapidity cuts corresponding to the fixed-target acceptance are imposed.

\begin{itemize}
    \item Boosting of the center-of-mass frame:
    \begin{align*}
        y_{\text{lab}} &= y + y_{\text{mid}}, \\
        y_{\text{mid}} &= -2.1,
    \end{align*}
    \item After boosting, pseudorapidity ($\eta$) is recalculated from $y$ using:
    \begin{equation}
        \eta = \frac{1}{2} \ln \left( \frac{\sqrt{m_T^2 \cosh^2 y - m^2} + m_T \sinh y}
        {\sqrt{m_T^2 \cosh^2 y - m^2} - m_T \sinh y} \right),
    \end{equation}
    where $m_T = \sqrt{m^2 + p_T^2}$ is the transverse mass.
\end{itemize}

This procedure allows us to directly compare collider-mode and fixed-target-mode results under equivalent kinematic selections. With this cross-check, we see that the measurement is independent of frame of reference. 

\begin{figure}[h!]
\includegraphics[width=\linewidth]{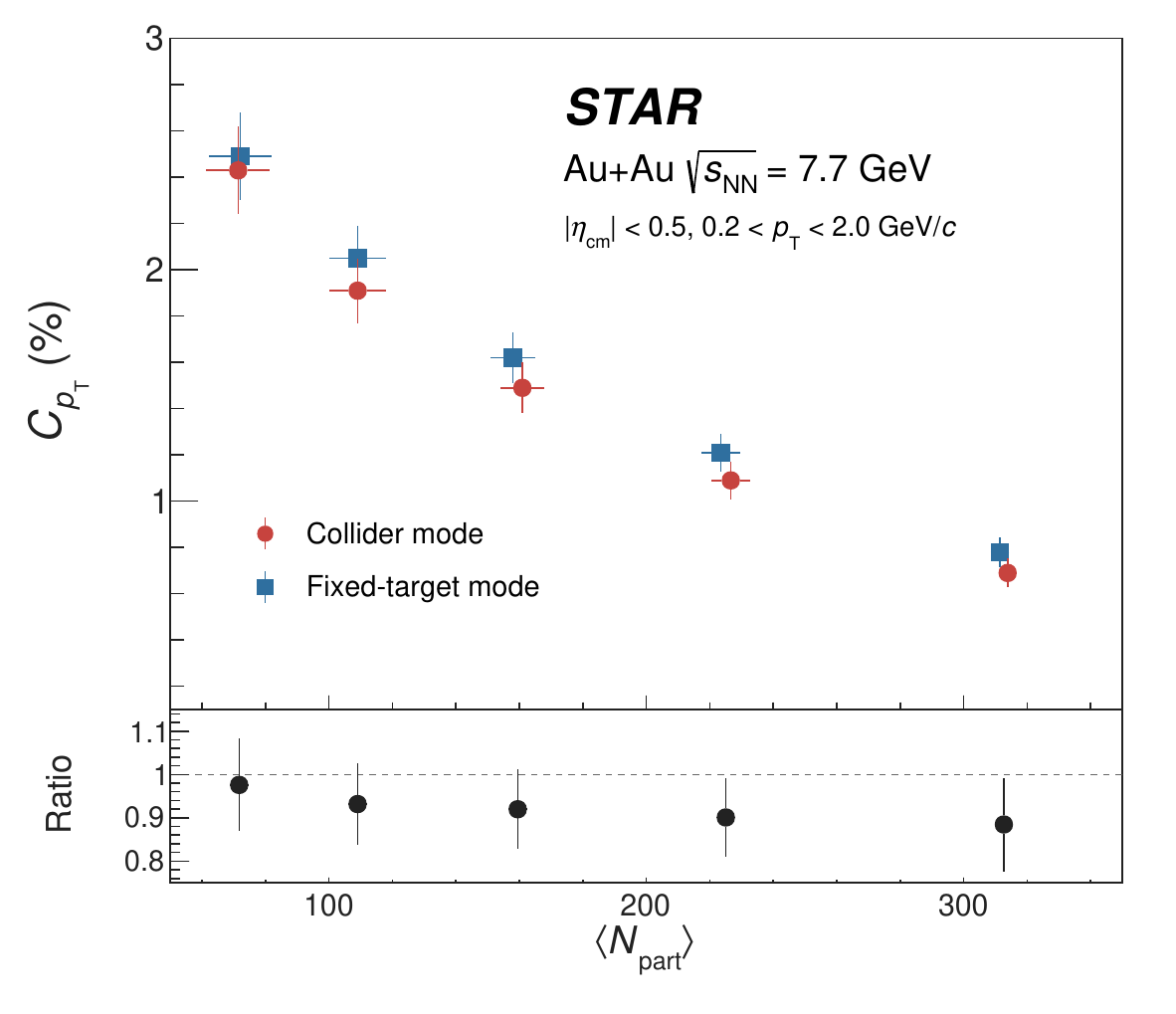}
\caption{\label{fig:S2}
The relative dynamical correlation $C_{p_{\rm T}}$ as a function of $\langle N_{\text{part}}\rangle$ for Au+Au collisions at $\sqrt{s_{\rm NN}} = 7.7$~GeV measured by STAR in collider and fixed-target modes. Statistical and systematic uncertainties are shown. The lower panel presents the ratio of collider-mode to fixed-target-mode results, demonstrating consistency between the two configurations within uncertainties.
}

\end{figure}

\subsection{Effect of Centrality Bin Width Effect}
\label{sec:supplemental-CBWC}

As a cross-check, we examine the effect of the centrality bin width correction (CBWC)
using collider-mode data at $\sqrt{s_{NN}} = 7.7$~GeV, where the reference multiplicity
RefMult is used for centrality determination. In the collider configuration, RefMult is
defined as the number of reconstructed primary charged-particle tracks within
$|\eta| < 1.0$.

\begin{figure}[h!]
\includegraphics[width=\linewidth]{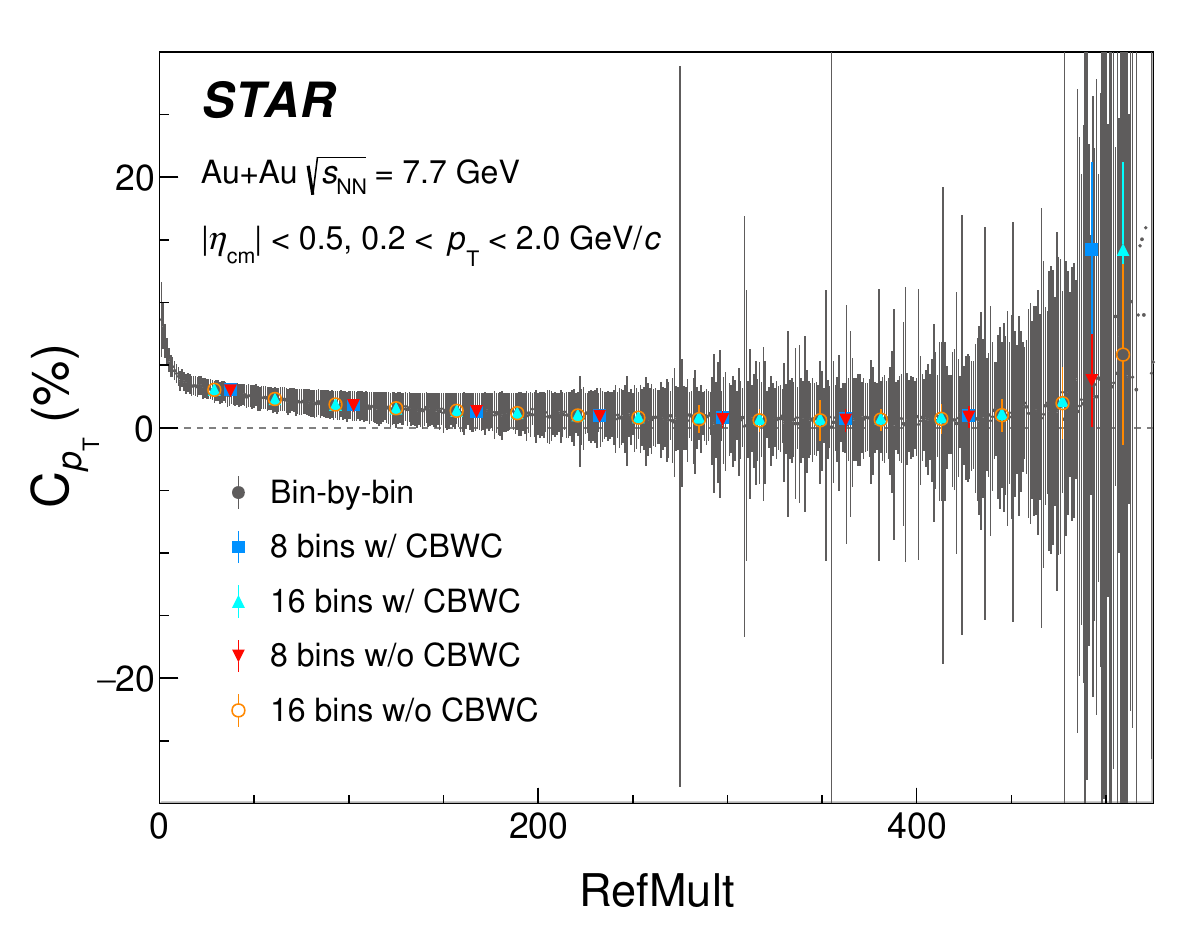}
\caption{\label{fig:S3}
The relative dynamical correlation $C_{p_{\rm T}}$ as a function of the reference
multiplicity (RefMult) for Au+Au collisions at $\sqrt{s_{NN}} = 7.7$~GeV in the collider
configuration.
The gray band shows the values of $C_{p_{\rm T}}$ obtained in unit RefMult bins, including
their statistical uncertainties.
Results corresponding to 8 and 16 centrality classes are shown, both with and without
applying the centrality-bin-width correction (CBWC).
Only statistical uncertainties are shown.
}
\end{figure}

An event-weighted averaging of observables over multiplicity bins, referred to as the
centrality bin width correction (CBWC), is performed when reporting results for a given
centrality class (e.g., 0--5\%). The CBWC accounts for variations of extensive quantities
across multiplicity bins within a centrality interval. Observables are calculated in each
multiplicity bin and then averaged using the number of events as weights, thereby removing
residual dependencies on the centrality-bin width~\cite{Luo_2013}.

We find that the quantity $C_{p_{\rm T}}$ is independent of the CBWC procedure and of the
chosen bin size, consistent with its intensive nature.

\subsection{Quantifying Non-monotonicity of measurements}
\label{sec:supplemental-quant}

\begin{figure}[h!]
\includegraphics[width=\linewidth]{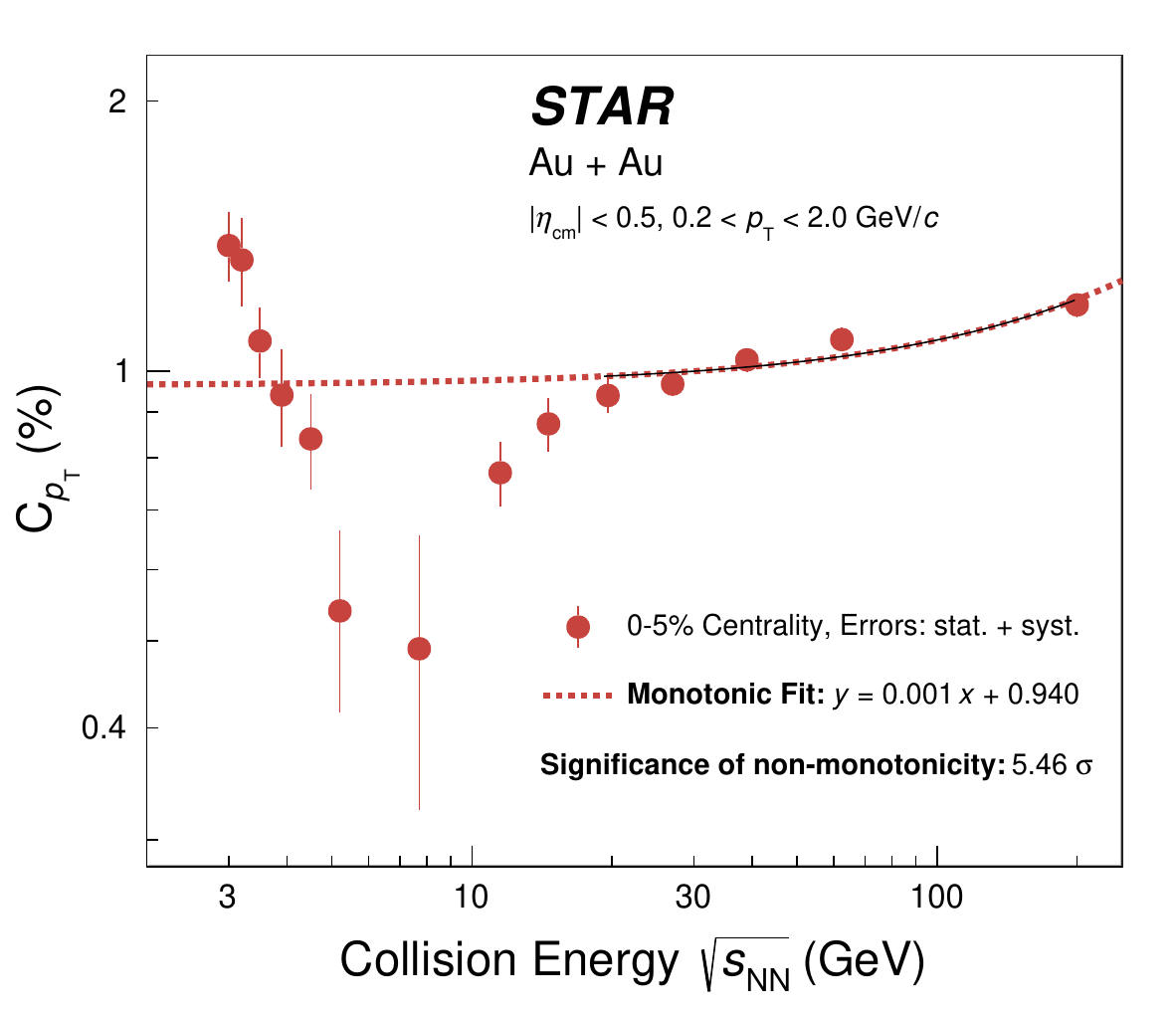}
\caption{\label{fig:S4} Non-monotonicity quantification for 0--5\% central Au+Au collisions from STAR at $\sqrt{s_\mathrm{NN}} = 3$--200~GeV. Statistical and systematic uncertainties are added in quadrature and shown as bars on the data points. The solid red line represents first-order polynomial fits to the data, with fit parameters and the significance of non-monotonicity displayed.}
\end{figure}

In this Letter we quote the significance of non-monotonicity by fitting the correlator's energy dependence to a strictly monotonic reference function, such as a first-order polynomial. To establish a robust, data-driven baseline and minimize model bias, the fit was anchored to high-energy data points with the smallest statistical uncertainties (down to $\sim20$~GeV), where the correlator's behavior is approximately monotonic. The deviation of the experimental data from this smooth, monotonic baseline provides an estimate, or an upper bound, on the significance of the observed non-monotonic structure, allowing direct assessment without overfitting statistical fluctuations. Furthermore, for the specific quantification at the $7.7\text{ GeV}$ collision energy, the final mean value was computed from both contributing measurements, and their associated uncertainties were combined in quadrature.

%

\end{document}